**Two-Dimensional Altermagnetism in Epitaxial CrSb Ultrathin Films**

*Keren Li, Yuzhong Hu, Yue Li, Ruohang Xu, Shaozhong Ma, Heping Li*, Kun Liu, Chen Liu, Lu Cao, Jincheng Zhuang, Yee Sin Ang, Jiaou Wang, Haifeng Feng*, Weichang Hao, Yi Du**

K. Li, Y. Hu, Y. Li, R. Xu, S. Ma, H. Li, L. Cao, J. Zhuang, H. Feng, W. Hao, Y. Du
School of Physics
Beihang University
Beijing 100191, P. R. China
E-mail: heping_li@buaa.edu.cn; haifengfeng@buaa.edu.cn; yi_du@buaa.edu.cn

K. Liu, C. Liu, J. Wang
Institute of High Energy Physics
Chinese Academy of Sciences
Beijing 100049, P. R. China

Y. Ang
Mathematics and Technology
Singapore University of Technology and Design
Singapore 487372, Singapore

Keren Li, Yuzhong Hu and Yue Li contributed equally to this work.

Keywords: altermagnetic, CrSb, ultrathin films, scanning tunneling microscopy, angle-resolved photoemission spectroscopy

Altermagnets constitute an emerging class of collinear magnets that exhibit zero net magnetization yet host spin-split electronic bands arising from non-relativistic spin-space-group symmetries. Realization of altermagnetism in the two-dimensional (2D) limit remains an outstanding challenge because dimensional reduction suppresses $k_Z$ dispersion and destabilizes the symmetry operations essential for spin compensation. Here, we investigate ultrathin CrSb films grown epitaxially on $Bi_2Te_3$ substrate and uncover the evolution of altermagnetism in the 2D limit. Scanning tunneling microscopy (STM), quasiparticle interference (QPI), angle-resolved photoemission spectroscopy (ARPES), and density functional theory (DFT) calculations show that interfacial symmetry breaking in the one-unit-cell (1 UC) limit gives rise to localized electronic states and uncompensated magnetic moments. These interfacial effects become weakened from 7/4 UC, accompanied by the recovery of a bulk-like coordination environment and the emergence of altermagnetic electronic characteristics. Our results show that the essential altermagnetic electronic structure of CrSb survives at a thickness of only ~1.05 nm, demonstrating the robustness of altermagnetism in the 2D limit and opening opportunities for integrating stray-field-free spin order into low dimensional spintronic architectures.

## 1. Introduction

Altermagnets represent an emerging class of collinear magnetic materials that, despite exhibiting zero net magnetization in real space like antiferromagnets, host non-relativistic spin-split electronic band structures in momentum space[1-6]. This counterintuitive behaviour originates from specific spin-space group (SSG) operations, including rotation, mirror, glide, or screw, which connect opposite spin sublattices in contrast to conventional operations like inversion ($P$) or translation ($t$)[7,8]. The symmetry operations generate alternating spin textures across the Brillouin zone (BZ), while preserving overall magnetic compensation[4]. The unique magnetic state of altermagnets combines the robustness and ultrafast dynamics of antiferromagnets with the spintronic functionality of ferromagnets, offering a stray-field-free platform for next-generation spintronic control[9,10]. Recently, altermagnetic spin splitting has been reported in several magnetic systems, including $RuO_2$[11-19], MnTe[20-24], $Mn_5Si_3$[25], and CrSb[26-32]. In these bulk materials, energy dispersion along the out-of-plane direction ($k_Z$ dispersion) plays a predominant role in shaping the altermagnetic spin texture. It is intimately related to specific SSG operations that enforce spin degeneracy along certain high-symmetry nodal surfaces. In altermagnetic CrSb, for instance, the spin degeneracy on the $k_Z = 0$ nodal surface is preserved because two spin-opposite sublattices are connected by non-relativistic SSG operations $[C_2||C_{6Z}t]$ or $[C_2||M_Z]$. Such the operations break the conventional spin group symmetry $[C_2||t]$ and space-time reversal symmetry $[T||P]$, producing an altermagnetic spin-splitting gap. The gap is further enhanced by Sb-mediated third-nearest-neighbour electron hopping along the Cr-Sb-Cr superexchange pathway[31]. Consequently, CrSb exhibits an unusually large non-relativistic spin splitting of approximately 1.0 eV, which is exceeds the thermal energy scale at room temperature.

In general, it is well accepted that the strong $k_Z$-dependence reflects the intrinsic reliance of altermagnetism on three-dimensional (3D) band topology, which is naturally satisfied in altermagnetic materials in bulk form. To meet the requests of spintronic-devices, however, it starves for realizing altermagnetism in two-dimensional (2D) regime, *i.e.* in ultrathin films. The dimensional constraint is primary not only to enable effective electronic gating and proximity coupling, but also to interfacial engineer and precisely modulate quantum states[33-35]. Achieving 2D altermagnets requires rational design of crystal architectures that inherently satisfy crystal symmetry, together with synthesis strategies capable of stabilizing the structures even at monolayer limit without introducing disorder or chemical degradation. Nevertheless, dimensional reduction would suppress $k_Z$ dispersion in addition to destabilize key crystal symmetries. As a result, altermagnetism may be weaken or even vanish in 2D regime, which limits altermagnetic applications in spintronics. Although, altermagnetism has been claimed recently in layered materials, such as $KV_2Se_2O$[36], $Rb_2V_2Te_2O$[37] and $CoNb_4Se_8$[38], the observed phenomena in fact arise in their bulk crystals where the interlayer coupling remains non-negligible. To date, the experimental realization of genuine 2D altermagnets has not been demonstrated yet, which possesses an urgent challenge of both profound scientific significance and technological importance.

Here, we realize 2D altermagnetism in unit-cell (UC) thin films derived from bulk altermagnet CrSb. Scanning tunneling microscopy (STM), scanning tunnelling spectroscopy (STS), quasiparticle interference (QPI), angle-resolved photoemission spectroscopy (ARPES) and density functional theory (DFT) calculations reveal that the interfacial effects dominating the 1 UC film are largely mitigated at 7/4 UC. In the 1 UC limit, localized electronic states and uncompensated magnetic moments emerge at the CrSb/$Bi_2Te_3$ interface, resulting in a ferrimagnetic-like interfacial state. With increasing thickness, the influence of the interface is reduced, and a bulk-like coordination environment is gradually restored, accompanied with the emergence of an altermagnetic electronic structure. These findings clarify how heterointerfaces influence altermagnetism in the 2D limit and provide a basis for controlling altermagnetic states in ultrathin films. This work paves the way for integrating altermagnetism into low dimensional spintronic devices and lays the theoretical and experimental groundwork for exploring 2D altermagnets.

## 2. Results and Discussion

CrSb crystallizes in a hexagonal NiAs-type structure with lattice constants $a = b = 4.12$ Å and $c = 5.89$ Å[39], where each Cr atom is octahedrally coordinated by six Sb atoms. As shown in **Figure 1a**, CrSb thin films with varying thicknesses were grown on $Bi_2Te_3$ substrates by molecular beam epitaxy (MBE, see **Methods** for details). The films adopt a single crystallographic orientation aligned with the substrate, as confirmed by STM and LEED measurements (**Figure S1**). The first terrace of CrSb films on the substrate has a height of 0.60 nm (**Figure 1a and 1d**), consistent with the thickness of 1 UC comprising four atomic layers. To reveal the growth dynamics, Cr and Sb were deposited on the $Bi_2Te_3$ substrates, respectively (**Figure S3**). It is found that deposition of Sb resulted in triangular islands with a height of ~0.5 nm. In contrast, deposition of Cr produced the Cr clusters which appeared as dark defects in STM images[40]. Such a fingerprint feature of Cr clusters was also observed in the exposed $Bi_2Te_3$ substrate regions after CrSb growth (**Figure 1b**), which indicates that the epitaxial growth is initiated by a Cr layer on $Bi_2Te_3$. Thus, epitaxial CrSb follows a Cr-Sb-Cr-Sb atomic stacking sequence in the case of 1 UC films.

With increasing film thickness, the terrace step height decreases to ~0.45 nm (**Figure 1d**), which is smaller than the thickness of 1 UC CrSb. Based on the crystal structure, such a height corresponds to the addition of three atomic layers (Cr-Sb-Cr or Sb-Cr-Sb), indicating an alternating three-atom-layer stacking mode. Since the interfacial layer is Cr-terminated, the terrace height can be expressed as $(1 + 3n)/4$ UC, where $n$ is a positive integer. This behaviour highlights the decisive role of the Cr-terminated interface in the subsequent epitaxial growth mode of CrSb films. As a result, CrSb films can be fabricated with different surface terminations such as 1 UC, 7/4 UC, and 10/4 UC. Atomic-resolution STM images and corresponding fast Fourier transform (FFT) patterns show an identical in-plane lattice constant of 0.44 nm for both substrate (**Figure 1b and 1e**) and 10/4 UC CrSb film (**Figure 1c and 1f**), which indicates that the lattices of CrSb films are slightly expanded to match the substrate.

The thickness-dependent terminations break the ideal bulk stacking symmetry, which creates two inequivalent Cr sites. As illustrated in **Figure 1g**, $Cr_1$ site locates in a trigonal pyramidal environment ($C_{3v}$), whereas $Cr_2$ site retains the bulk-like octahedral coordination ($O_h$). The different crystal field environments lead to distinct orbital splitting[41], as shown in **Figure 1h**. Crystal-field analysis predicts three unpaired electrons for both $Cr_1^{3+}$ and $Cr_2^{3+}$. But the reduced coordination and lower symmetry of $Cr_1$ sites enhance electron localization and increase the orbital contribution to its magnetic moment. This effect is expected to result in a larger local magnetic moment at $Cr_1^{3+}$ site than at $Cr_2^{3+}$ site.

To probe these thickness-dependent electronic states, STS measurements were performed on CrSb films of varying thicknesses. Owing to the distinct surface terminations of the 1 UC (Sb-terminated) and 7/4 UC (Cr-terminated) films, an atomic-scale lateral displacement is expected between the adjacent terraces (**Figure 2a**). This feature is clearly identified in the atomic-resolution STM image in **Figure 2c**. Four $dI/dV$ spectra were acquired at representative positions along the dashed line across the terrace boundary in **Figure 2b**. As depicted in **Figure 2d**, the spectra reveal distinct differences among the 1 UC, 7/4 UC (upper panel), and substrate (lower panel). The substrate exhibits a lower density of states (DOS) near the Fermi level ($E_F$), distinguishing it from the metallic CrSb films (upper panel). Both 1 UC and 7/4 UC films exhibit two peaks in the energy range of -0.2 V to -0.4 V, denoted as $P_2$ and $P_3$ peaks (**Figure 2d and 2e**). In contrast, an additional peak ($P_1$ peak) emerges at approximately -0.065 V exclusively in the 1 UC film. This peak can be well fitted by an asymmetric Fano resonance line shape[42] (detailed in **Figure S4**). Such characteristic asymmetric indicates quantum interference between a localized discrete state and an itinerant continuum, suggesting the presence of a localized electronic state in the 1 UC film. To further clarify its origin, we performed STS measurements across a defect in the 1 UC film (**Figure 2g**). The spectra exhibit a clear band bending (**Figure 2h**), consistent with negatively charged Cr-related defects. Meanwhile, the $P_1$ peak persists in defect-free regions while being absent at the defect centre (**Figure S5**). This observation excludes the defect-induced origin and instead identifies $P_1$ peak as an intrinsic feature of the 1 UC film and. Therefore, we attribute the emergence of this localized state ($P_1$ peak) to symmetry breaking within the 1 UC film.

In fact, we note that the interfacial $Cr_1$ layer may further experience crystal-field modification from the substrate, which breaks the specific rotation ($[C_2||C_{6Z}t]$) and mirror ($[C_2||M_Z]$) SSG operations that connect the two spin-opposite Cr sublattices in the bulk. Such symmetry breaking disrupts the fundamental conditions for altermagnetism and likely induces unequal magnetic moments between $Cr_1^{3+}$ and $Cr_2^{3+}$, generating local uncompensated moments. From this perspective, the $P_1$ peak is closely related to the interfacial symmetry breaking.

As the thickness increases to 7/4 UC, the disappearance of the $P_1$ peak suggests a reduced interfacial effect. The additional Sb-Cr-Sb stacking produces a bulk-like coordination environment and partially restores the local symmetry connecting the two Cr sublattices (**Figure 1j**). Furthermore, $dI/dV$ spectra obtained from 10/4 UC and 13/4 UC CrSb films (**Figure 2f**) exhibit identical features, which is consistent with bulk CrSb. These results indicate that a bulk-like electronic structure is established at thicknesses beyond 7/4 UC.

To further visualize the interfacial electronic structure, QPI mapping was employed on the 1 UC CrSb/$Bi_2Te_3$. **Figure 3a** shows a representative STM topography, and the corresponding spatial map of the local density of states (LDOS) at $V_s$ = 0.3 V is presented in **Figure 3b**. The Fourier-transformed QPI (FT-QPI) patterns at various energies are shown in **Figure 3c-f** (six-fold symmetrized data in **Figure S6**). The energy-momentum dispersion extracted from the six-fold symmetrized FT-QPI patterns remains a linear tendency (**Figure S7a**). This rules out scattering from trivial parabolic bands of the CrSb film or $Bi_2Te_3$ bulk states, suggesting that the observed QPI signals are primarily associated with interfacial states that largely retain the Dirac-like characteristics of the topological surface states (TSS). The FT-QPI patterns exhibit energy-dependent evolution. At higher energies away from the Dirac point ($V_s$ = 0.2 V and 0.3 V, **Figure 3e-f**), the scattering is dominated by vectors along the $\overline{\Gamma}$-$\overline{M}$ direction ($\boldsymbol{q}_{\overline{\Gamma}\text{-}\overline{M}}$). As the energy approaches the Dirac point ($V_s$ = 0.1 V and 0.15 V, **Figure 3c-d**), scattering vectors along the $\overline{\Gamma}$-$\overline{K}$ direction ($\boldsymbol{q}_{\overline{\Gamma}\text{-}\overline{K}}$) emerge. This evolution is closely related to the geometry of the constant-energy contours (CECs) of the TSS[43] (**Figure S7b-d**). At higher energies, the CECs exhibit strong hexagonal warping. This warping introduces a significant phase mismatch and out-of-plane spin canting, which suppresses the $\overline{\Gamma}$-$\overline{K}$ scattering. As the energy moves closer to the Dirac point, the hexagonal warping weakens, the CEC becomes more isotropic, and the $\boldsymbol{q}_{\overline{\Gamma}\text{-}\overline{K}}$ scattering becomes clearly observable.

However, the relaxation of hexagonal warping alone cannot fully explain the appearance of the $\boldsymbol{q}_{\overline{\Gamma}\text{-}\overline{K}}$ scattering. Comparable $\boldsymbol{q}_{\overline{\Gamma}\text{-}\overline{K}}$ scattering is absent on the exposed $Bi_2Te_3$ substrate (**Figure S8**), which indicates an additional scattering channel is introduced after the growth of 1 UC CrSb. Since backscattering along the $\overline{\Gamma}$-$\overline{K}$ direction is strictly forbidden in pristine $Bi_2Te_3$ by the spin-momentum locking of the TSS[44], purely structural defects cannot induce spin-flip scattering. Thus, the emergence of $\overline{\Gamma}$-$\overline{K}$ scattering in the 1 UC CrSb film on $Bi_2Te_3$ indicates a breakdown of this time-reversal symmetry. Combined with our STS analysis above, this strongly suggests that time-reversal symmetry is broken by uncompensated magnetic moments at the interface.

ARPES measurements further corroborate the interfacial electronic modification. As shown in **Figure 3g**, the pristine $Bi_2Te_3$ spectra exhibit a well-resolved Dirac surface state together with the electron pocket of the bulk conduction band (BCB) and the characteristic M-shaped bulk valence band (BVB) along the $\overline{\Gamma}$-$\overline{K}$ direction. The momentum distribution curves (MDCs) track this pristine TSS with red squares. After the deposition of 1 UC CrSb (**Figure 3h**), the energy positions of the BCB and BVB remain nearly unchanged, which excludes substantial interfacial charge transfer. In contrast, the original TSS becomes indistinct, and an additional dispersive band emerges, highlighted by the blue triangles in the MDCs. The corresponding energy distribution curves (EDCs) further support this in **Figure 3i-j**. The red arrow marks the intrinsic band of pristine $Bi_2Te_3$, whereas a new band appears after the growth of 1 UC CrSb (blue arrow). This newly emerged feature is consistent with the dispersive band identified from the MDC analysis.

The modification of the TSS, strongly suggests that the influence of the CrSb layers is primarily localized at the interface[45,46]. Such behaviour is consistent with an interfacial reconstruction of the surface electronic structure arising from hybridization between CrSb and $Bi_2Te_3$ derived states. However, hybridization alone is insufficient to account for the appearance of $\boldsymbol{q}_{\overline{\Gamma}\text{-}\overline{K}}$ observed in QPI. Together with the QPI results, the ARPES data suggest that both interfacial hybridization and magnetic interactions associated with the uncompensated interfacial magnetic moments contribute to the electronic reconstruction at the CrSb/$Bi_2Te_3$ interface[47].

While STS measurements indicate that the local interfacial effects are weakened at 7/4 UC

and a bulk-like electronic structure gradually emerges with increasing thickness, the momentum-resolved evolution of the electronic structure remains unclear. **Figure 4a** and **4c** show the ARPES spectra of the 7/4 UC CrSb/$Bi_2Te_3$ acquired along $\overline{\Gamma}$-$\overline{M}$ and $\overline{\Gamma}$-$\overline{K}$ directions, respectively. Compared with the 1 UC CrSb film (**Figure 3h**), the 7/4 UC CrSb film exhibits two well-defined dispersive bands (blue and orange triangles) across $E_F$, indicating that the intrinsic electronic structure of CrSb is already established at this thickness. The corresponding MDCs and EDCs further resolve these bands (**Figure 4b** and **4d**), with the red squares marking the $Bi_2Te_3$ substrate bands and the blue and orange triangles tracing the two bands from CrSb. Along the $\overline{\Gamma}$-$\overline{M}$ direction, the two bands remain well separated in momentum, whereas along the $\overline{\Gamma}$-$\overline{K}$ direction they become much closer, revealing a pronounced in-plane anisotropy of the 7/4 UC CrSb electronic structure.

To explore the origin of these spectral features, DFT calculations were performed for bulk CrSb and relaxed CrSb/$Bi_2Te_3$ heterostructures with different CrSb thicknesses. Additional calculated results and analyses are provided in the Supporting Information. Unlike previous studies on thicker CrSb films approaching the bulk limit[32], the present work focuses on the ultrathin CrSb/$Bi_2Te_3$ heterostructure regime, where quantum confinement and interfacial coupling strongly modify the electronic structure. As shown in **Figure 4f**, bulk CrSb exhibits a characteristic $k_Z$-dependent band feature near $E_F$, which serves as a distinct electronic fingerprint of the altermagnetic phase. The CrSb-projected band structures of the CrSb/$Bi_2Te_3$ heterostructures reveal a clear thickness evolution (**Figure 4g-i**). In the 1 UC film, the characteristic band is absent because the electronic states near $E_F$ are strongly modified by quantum confinements and interface effects. At 7/4 UC, however, the characteristic band emerges near $E_F$ that closely resembles the bulk electronic fingerprint. This feature persists in thicker films (from 10/4 UC to 16/4 UC, **Figure4i and Figure S12**), and it is gradually consistent with the bulk. The emergence of this band feature at 7/4 UC indicates that the characteristic electronic fingerprint of bulk CrSb is already established at a thickness of only ~1.05 nm. Therefore, despite remaining in the 2D limit, 7/4 UC CrSb/$Bi_2Te_3$ already hosts the essential altermagnetic electronic structure of bulk CrSb, demonstrating the robustness of the altermagnetic electronic state in the ultrathin films.

## 3. Summary

In summary, we investigated the evolution of altermagnetism in ultrathin CrSb films epitaxially grown on $Bi_2Te_3$. At the 1 UC CrSb limit, interfacial coupling reconstructs the electronic structure and gives rise to a ferrimagnetic-like interfacial state. With increasing CrSb thickness, the interfacial effect is progressively suppressed, while the intrinsic electronic structure of CrSb is restored. The characteristic electronic fingerprint of altermagnetic CrSb already emerges at a thickness of only 7/4 UC (~1.05 nm), demonstrating that the essential altermagnetic electronic structure can be established well within the 2D limit. These findings clarify the interplay between quantum confinement, interfacial coupling, and the emergence of altermagnetism, providing important insights into the dimensional evolution of altermagnets. Our work establishes 7/4 UC as the critical thickness for recovering the intrinsic altermagnetic electronic structure, making the CrSb/$Bi_2Te_3$ heterostructure an ideal platform for probing low-dimensional altermagnetism. This work paves a viable route for manipulating altermagnetic states through interface engineering, laying a crucial foundation for next-generation low-dimensional spintronic devices.

## 4. Methods

*MBE growth of CrSb films on $Bi_2Te_3$*: The CrSb films were grown on $Bi_2Te_3$ substrates by MBE in an ultrahigh vacuum chamber (base vacuum $1\times10^{-10}$ mbar). After the substrate was heated to the growth temperature of 200 °C, high-purity chromium (Cr) and antimony (Sb) were evaporated using a resistively heated tantalum boat-type thermal evaporator and a standard Knudsen cell, respectively. The flux ratio of Cr:Sb was approximately 1:10.

*STM measurements and analysis*: All STM measurements were performed in ultrahigh vacuum (base vacuum $1\times10^{-10}$ mbar) at 77 K (Quantum Scale, NanoTech STM). A tungsten tip was used. All STM images were acquired in the constant-current mode. The d$I$/d$V$ spectra of CrSb were acquired at 77 K using standard lock-in detection, with a modulation frequency of 973 Hz and a modulation amplitude of 1~10 mV superimposed on the sample bias. WSxM software was used to process all STM images[48].

*ARPES measurements*: ARPES electronic band structure characterizations were performed using an Omicron-Scienta ARPES system with a He-I ($h\nu$ = 21.2 eV) light source at 10 K, detected by DA-30L analyzer. The epitaxial thin films were transported to the ARPES system via a vacuum suitcase. All measurements were performed under a base pressure better than $6\times10^{-11}$ mbar, and the vacuum of suitcase was maintained better below $1\times10^{-9}$ mbar.

*DFT calculations*: All calculations were performed based on the density functional theory (DFT), as implemented in the Vienna Ab initio Simulation Package (VASP)[49]. The Perdew-Burke-Ernzerhof (PBE) exchange-correlation functional was used within the plane-wave basis set and the projector augmented-wave (PAW) method[50,51]. The plane-wave energy cutoff was set to 360 eV. The generalized gradient approximation plus U (GGA + U) method are adopted, where U = 3.5eV is used for the Cr atom. Γ-centered k-point meshes of 13 × 13 × 6 and 9 × 9 × 1 were used for structural relaxation of the bulk and monolayer structures, respectively. For electronic structure calculations, denser Γ-centered k-point meshes of 15 × 15 × 12 (bulk) and 14 × 14 × 1 (monolayer) were adopted. Employing and fixing the experimental lattice symmetries, the crystal structure of CrSb was fully relaxed for structural relaxation until the atomic forces were smaller than 0.01 eV/Å on each atomic site. The electronic iteration was performed until the total energy change was smaller than $10^{-6}$ eV. The processing of Fermi surface also used VASPKIT and FermiSurfer[52,53]. The lattice parameters are a = b = 4.418 Å, corresponding to the experimental values.

**Data availability**

All data are available in the main text or the supplementary materials. Further data are available from the corresponding author on reasonable request.

**Acknowledgements**

We are grateful to the Analysis & Testing Center of Beihang University for the facilities and the scientific and technical assistance. This work was supported by the National Key R&D Program of China (2022YFB3403400), the Fundamental Research Funds for the Central Universities (501XSKC2025119001), the National Natural Science Foundation of China (12274016, 52473287, 12404196) and China Postdoctoral Science Foundation (2025M774220).

**Conflict of Interest**

The authors declare no conflict of interest.

**Author contributions**

Y. D. initiated and designed the overall research framework. Y. D., H. F., and H. L. provided conceptualization, methodology, data curation, validation, funding acquisition, reviewing, and editing with input from all coauthors. K. L. performed single crystal growth, STM characterizations, data analysis, and the original draft writing. Y. H. and R. X. performed ARPES characterizations, data analysis, and the original draft writing. S. M performed the XPS measurements. K. L. assisted in the ARPES measurements. Y. L., H. L. carried out DFT calculations. W. C. H., J. W., Y. S. A., J. Z., L.C. and C. L. provide valuable discussion.

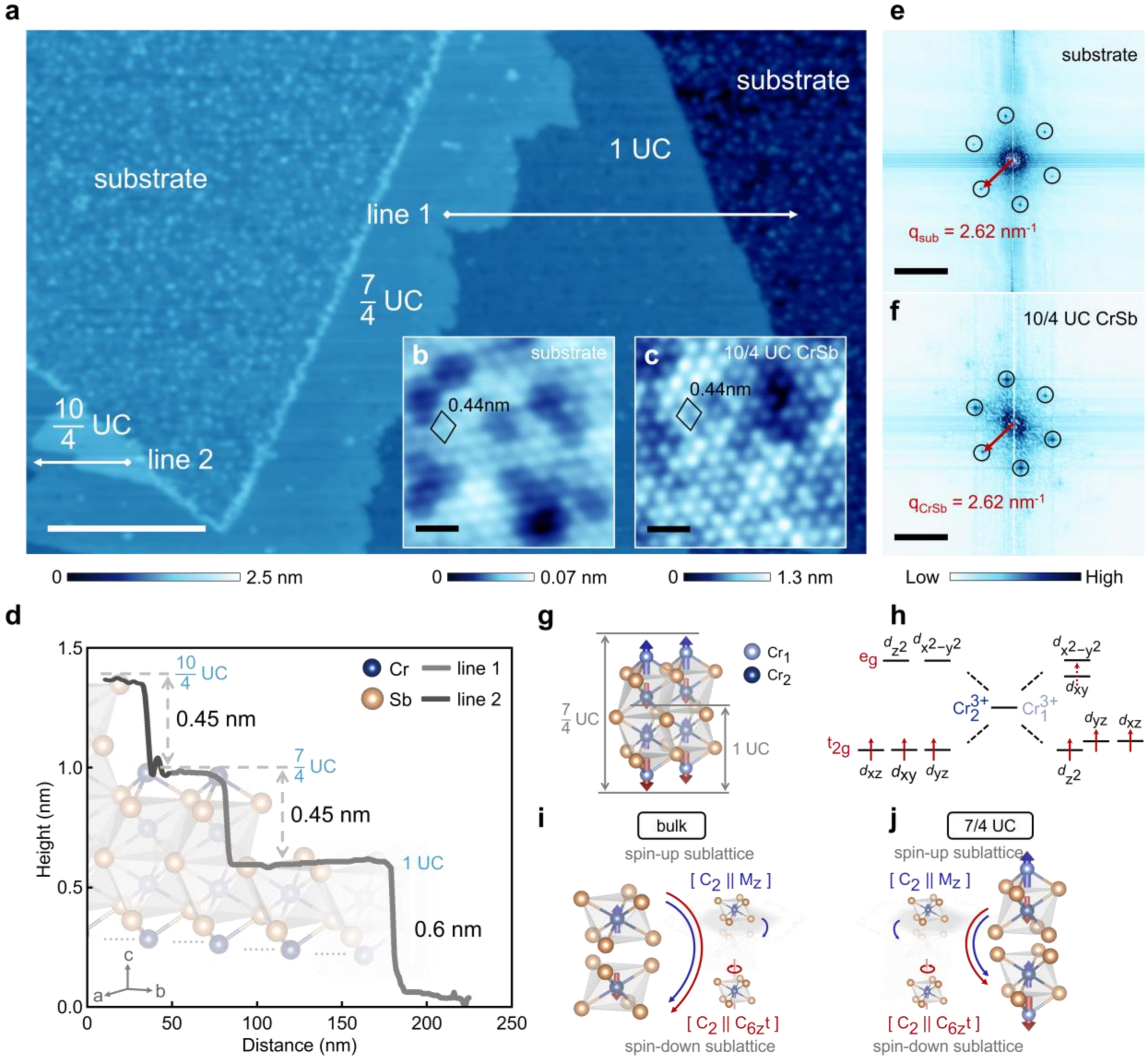

**Figure 1 | Crystal structure and STM characterization of CrSb atomic layers grown on $Bi_2Te_3$. a,** Large-scale STM topography of CrSb films grown on $Bi_2Te_3$ ($V$ = 5 V, $I$ = 10 pA). Scale bar, 80 nm. **b,** Atomic-resolution STM image of an exposed substrate $Bi_2Te_3$ after CrSb growth ($V$ = 1V, $I$ = 100 pA), showing Cr-related antisite defects. Scale bar, 1 nm. **c,** Atomic-resolution STM image of 10/4 UC CrSb ($V$ = -0.1V, $I$ = 100 pA). Scale bar, 1 nm. **d,** Height profile along the white lines in **a**, confirming quantized step heights. **e, f,** Fast Fourier-transformed patterns of **b** (exposed $Bi_2Te_3$ substrate) and **c** (10/4 UC CrSb), respectively. Scale bar, 3.2 nm$^{-1}$. **g,** Crystal structure schematic of CrSb grown on $Bi_2Te_3$. **h,** Crystal field splitting analysis of Cr 3$d$ orbital energies in octahedral ($O_h$, Cr2) and trigonal ($C_{3v}$, Cr1) coordination environments. **i, j,** Spin-sublattice symmetry operations in bulk CrSb and 7/4 UC CrSb, respectively.

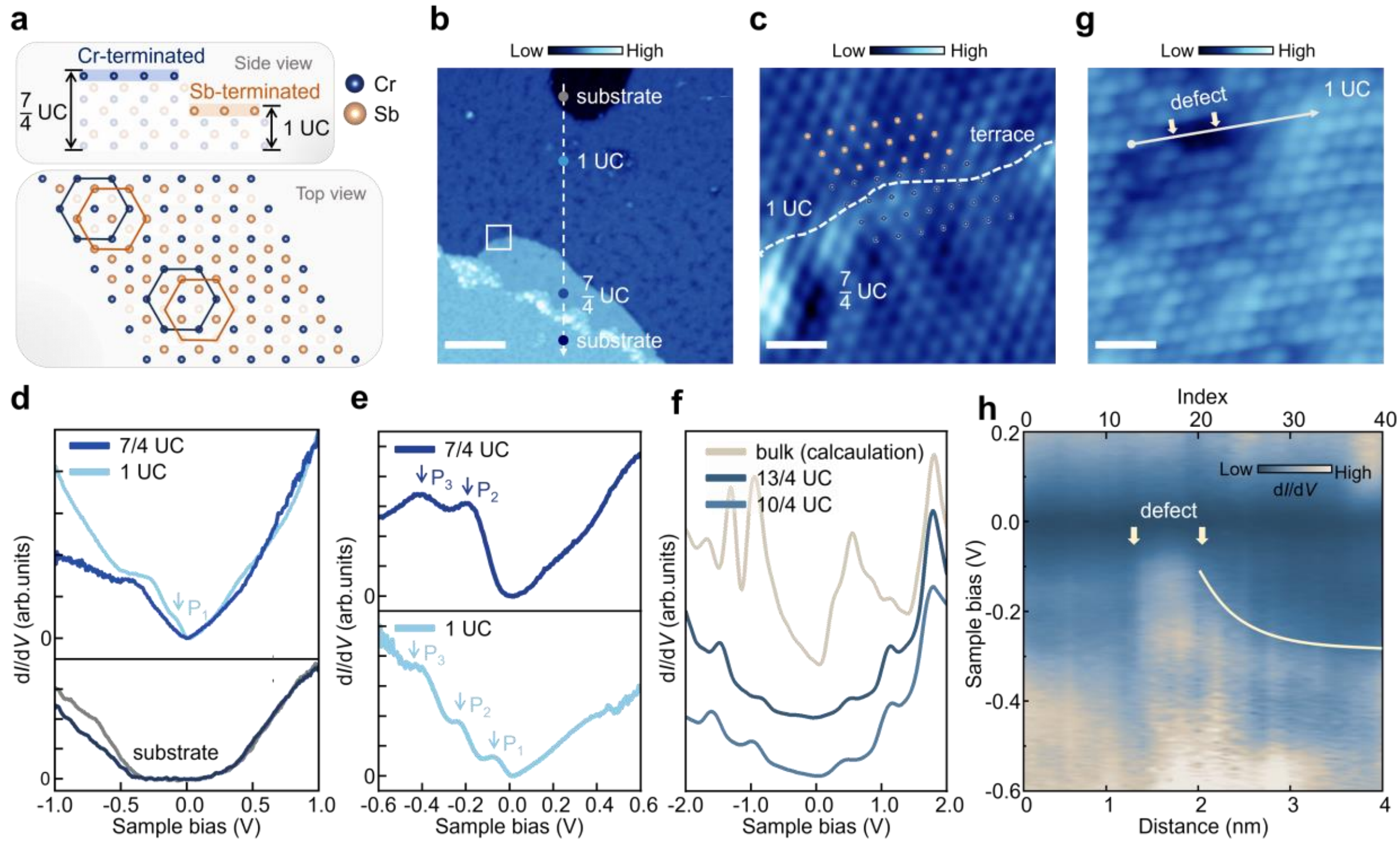

**Figure 2 | STS characterizations of varying thicknesses CrSb films grown on $Bi_2Te_3$. a,** Side view and top view structural models of the Sb-terminated 1 UC and Cr-terminated 7/4 UC CrSb films, highlighting the atomic lateral shift expected between the two surface terminations. **b,** Large-scale STM image of different thicknesses CrSb films grown on $Bi_2Te_3$ ($V$ = 2V, $I$ = 10 pA). Scale bar, 20 nm. **c,** Atomic-resolution STM image of the region marked by the white box in **b**, revealing the lateral displacement of the atomic lattice across the terrace boundary ($V$ = 0.8 V, $I$ = 400 pA). Scale bar, 1.2 nm. **d,** d$I$/d$V$ spectra acquired at the four representative points along the white dashed line in **b**, crossing from the $Bi_2Te_3$ substrate to the CrSb films. **e,** Comparison of d$I$/d$V$ spectra measured on the 1 UC and 7/4 UC CrSb, respectively. An additional resonance feature ($P_1$ peak) appears exclusively in the 1 UC film, whereas $P_2$ and $P_3$ peaks are present in both thicknesses. **f,** d$I$/d$V$ spectra acquired on the 10/4 UC and 13/4 UC CrSb films, with the calculated DOS of bulk CrSb, showing the emergence of a bulk-like electronic structure. **g,** Atomic-resolution STM image of 1 UC CrSb around defects. ($V$ = 0.2 V, $I$ = 400 pA). Scale bar, 1.2 nm. **h,** Waterfall image of 40 successive d$I$/d$V$ curves measured along the golden line in **g**, showing the band bending around the defect.

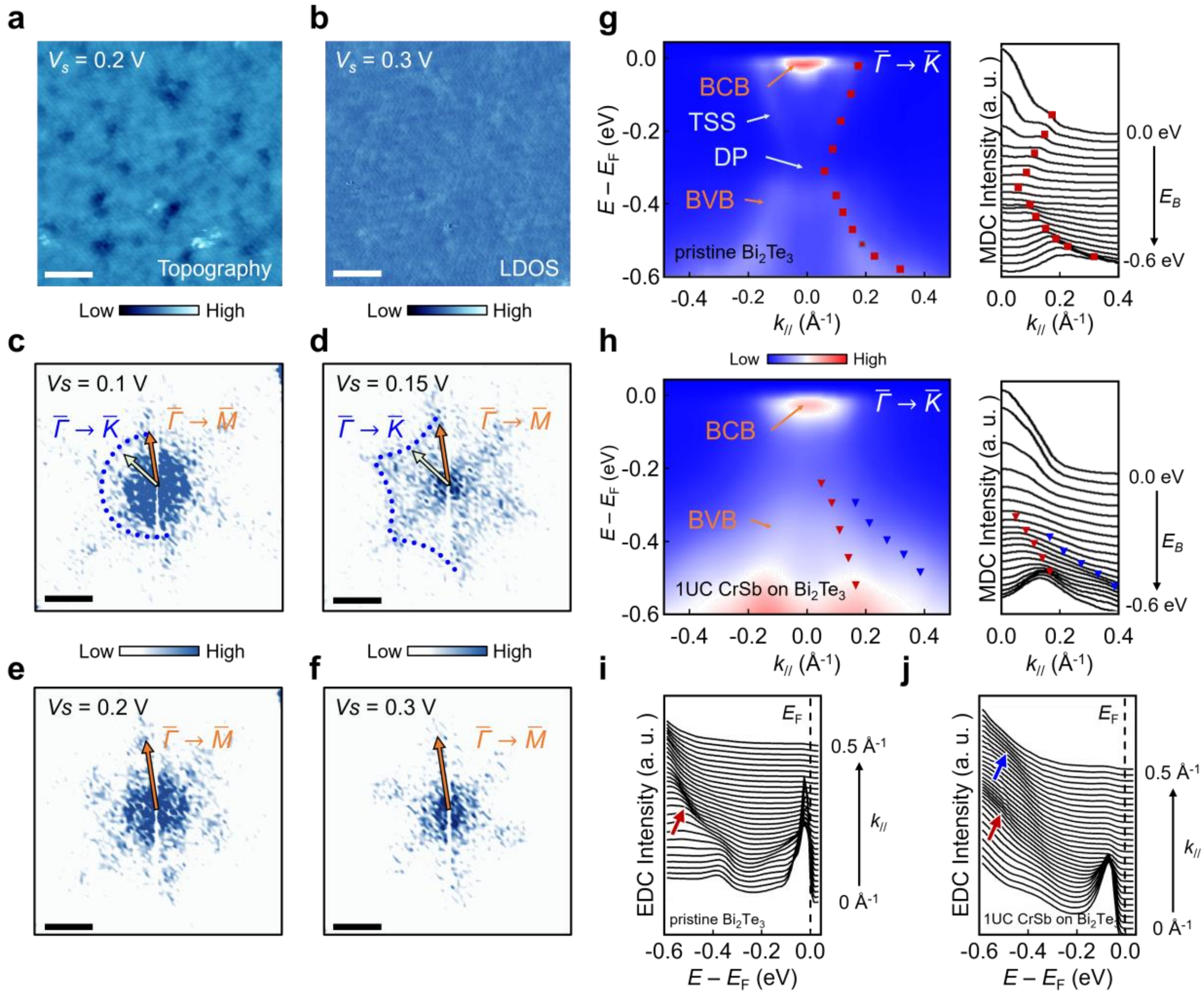

**Figure 3 | Quasiparticle interference (QPI) mapping and ARPES measurements of 1 UC CrSb film grown on $Bi_2Te_3$. a,** STM topography of 1 UC CrSb on $Bi_2Te_3$ ($20\times 20$ nm$^2$, $V = 0.2$ V, $I = 100$ pA). Scale bar, 4 nm. **b,** Spatial maps of the local density of states (LDOS). Scale bar, 4 nm. **c-f,** Corresponding Fourier-transformed QPI (FT-QPI) maps acquired at the indicated voltages. Scale bar, 0.8 nm$^{-1}$. The orange and green arrows indicate the $\bar{\Gamma}$-$\bar{M}$ and $\bar{\Gamma}$-$\bar{K}$ directions, respectively. The blue dots represent the scattering along the $\bar{\Gamma}$-$\bar{K}$ direction. **g-h,** ARPES spectra and corresponding momentum distribution curves (MDCs) of the substrate and 1 UC CrSb on $Bi_2Te_3$ film along $\bar{\Gamma}$-$\bar{K}$ direction, respectively. **i-j**, Corresponding energy distribution curves (EDCs), the red arrow represents the features from pristine $Bi_2Te_3$ substrate and the blue one represents an additional band after the growth of 1 UC CrSb.

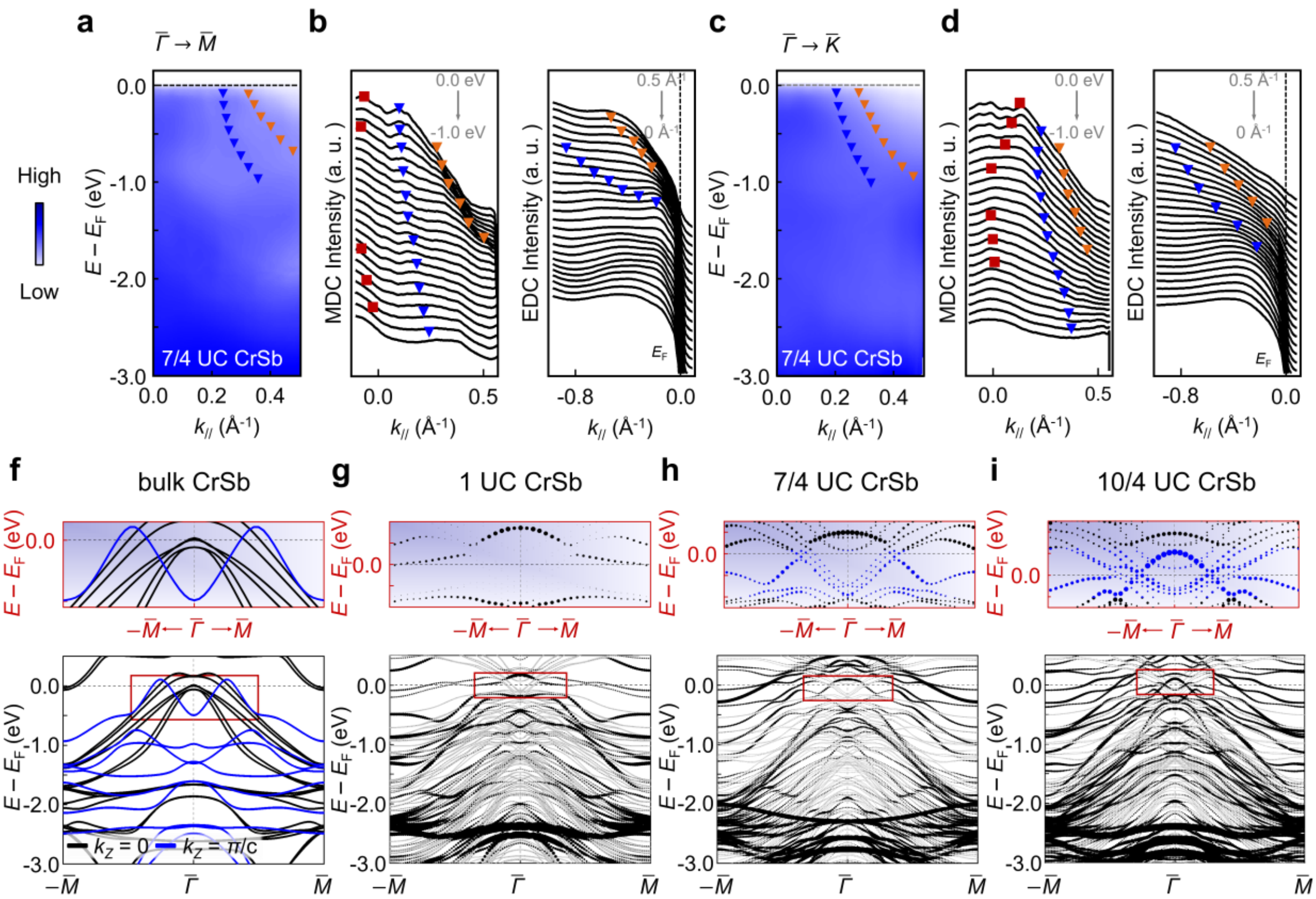

**Figure 4 | Thickness-dependent evolution of the electronic structure in CrSb films**. **a,** ARPES dispersion plot of 7/4 UC CrSb film on $Bi_2Te_3$ along the $\overline{\Gamma}$-$\overline{M}$ direction, the blue and orange triangles serve as guides to the eye. **b,** Corresponding MDCs (left), and EDCs (right), extracted near $\overline{\Gamma}$ point along the $\overline{\Gamma}$-$\overline{M}$ direction. Red squares trace the substrate bands and the blue and orange triangles trace the CrSb dispersive bands. **c,** APRES dispersion plot of 7/4 UC CrSb film on $Bi_2Te_3$ along the $\overline{\Gamma}$-$\overline{K}$ direction. **d,** Corresponding MDCs (left), and EDCs (right), extracted near $\overline{\Gamma}$ point along the $\overline{\Gamma}$-$\overline{K}$ direction. **f,** Calculated band structure of bulk CrSb. Black and blue curves correspond to the $k_Z = 0$ and $k_Z = \pi/c$ planes, respectively. The red box highlights a dispersive band near the Fermi level ($E_F$) that exists only in the $k_Z = \pi/c$ plane. The upper panel shows an enlarged view of the red-boxed region. **g-i**, Projected band structures of CrSb/$Bi_2Te_3$ heterostructures with CrSb thicknesses of 1 UC, 7/4 UC, and 10/4 UC, respectively. The upper panels display enlarged views of the red-boxed regions in the corresponding lower panels. The solid grey lines represent the total bands structure of the heterostructure, while the black marks represent the states projected onto the CrSb layers. The highlighted region contains a characteristic band feature (blue dots) near the $\overline{\Gamma}$ point and close to $E_F$. This feature is absent in the 1 UC film, appears at 7/4 UC, and remains visible with further increasing thickness.